\documentstyle[aps,prd,amssymb,eqsecnum,epsf,twocolumn]{revtex}
\newcommand{\be}{\begin{equation}}
\newcommand{\ee}{\end{equation}}
\newcommand{\bea}{\begin{eqnarray}}
\newcommand{\eea}{\end{eqnarray}}
\newcommand{\nn}{\nonumber}

\newcommand{\newc}{\newcommand}
\newc{\ra}{\rightarrow}
\newc{\lra}{\leftrightarrow}
\newc{\beq}{\begin{equation}}
\newc{\eeq}{\end{equation}}
\newc{\ba}{\begin{eqnarray}}
\newc{\ea}{\end{eqnarray}}
\newskip\humongous \humongous=0pt plus 1000pt minus 1000pt

\newif\ifdtup

\relax
\title{Quadrupole Instabilities of Relativistic Rotating Membranes}
\author{M. Axenides$^a$, E. G. Floratos$^a$,$^b$, L. Perivolaropoulos$^a$
} 
\address{$^a$ Institute of Nuclear Physics, N.C.S.R. Demokritos,  153
10, Athens, Greece, \\e-mail:axenides@mail.demokritos.gr, 
leandros@mail.demokritos.gr, http://leandros.chem.demokritos.gr 
\\ $^b$ National University of Athens, Athens, Greece, e-mail:manolis@mail.demokritos.gr} 

\date{\today}
\draft

\begin{document}

\maketitle
\begin{abstract}
We generalize recent study of the stability of isotropic
(spherical) rotating membranes to the anisotropic ellipsoidal
membrane. We find that while the stability persists for
deformations of spin $l=1$, the quadrupole and higher spin 
deformations ($l\geq 2$) lead to instabilities. We find the 
relevant instability modes and the corresponding eigenvalues. 
These indicate that the ellipsoidal rotating membranes generically 
decay into finger-like configurations. 
\end{abstract}


\section{Introduction}
%The best candidate for the unification of the weak and strong
%sectors of string theories is currently the M-theory\cite{TWSD}
%which can be framed together with its ingredients into Matrix
%theory\cite{BFSS}. The conjecture of this theory is that all the
%degrees of freedom of various sectors of the five known string
%theories can be represented by appropriate operators of the Matrix
%theory using duality properties. Up to now all perturbative checks
%of this idea although not straight forward have
% been successful and focus on the nonperturbative sector leads to the
%study of classical solutions of Matrix theory in general
%backgrounds\cite{t01}. Recent progress in this direction is the
%successful formulation of Matrix theory in weak gravitational and
%gauge backgrounds i.e. dynamics of matrix branes in the background
%of their mutual and external forces\cite{RCM}.

The eleven dimensional classical and quantum supermembrane is one
of the poorly understood elements of M-Theory\cite{TWSD,BFSS}. The
study of this sector is important both for the understanding of
the strongly coupled string theories as well as for the
quantization of the supermembrane. Recent interest for the
classical solutions of the Matrix theory representing $D_0$-branes
attached to spherical membranes is explained as a first step to a
formulation of Matrix theory in weak external gravitational and
gauge backgrounds \cite{t01,RCM,KT}. Particular solutions of the
classical matrix equations representing rotating ellipsoidal
configurations of N $D_0$ branes attached to a membrane which
exhibit stability properties have been proposed and their
semiclassical spectrum has been studied \cite{HS}.

In a recent paper\cite{afp00} we found that the isotropic
(spherical) rotating membrane is stable under all modes of small
multipole deformations and we found explicitly the corresponding
spectrum and the eigenmodes. Here we generalize that study by
finding in detail the stability properties of the rotating {\it
ellipsoidal} membrane which is an inhomogeneous solution of the
bosonic part of the supermembrane equations restricted to six
spatial dimensions. We find that while the stability persists for
deformations of spin $l=1$, the higher spin deformations ($l\geq
2$) lead to instabilities. We find the relevant modes and the
corresponding eigenvalues.

It is well known that spherical matrix and membrane solutions are
isomorphic \cite{KT,afp00}. Moreover the linearized problems for
the fluctuations preserve the same isomorphism due to the specific
spin $1$ form of the solution. The matrix solution is a bound
state of N $D_{0}$ branes attached on a $D_{2}$ brane whose
stability properties is obtained using our continuous membrane
investigation. Indeed one only has to replace the perturbations
$\delta X_i \;=\; \sum_{m} \epsilon_{i}^{m} Y_{lm}$ by the matrix
fluctuations $ \hat{ \delta X}_i \;=\; \sum _{m} \epsilon ^{m}_{i}
\hat {Y}_{lm} $ where $\hat{Y}_{lm}$ are the $SU(2)$ tensor
spherical harmonics\cite{afp00}.

The isomorphism with the stability analysis of matrix solutions
persists also in the anisotropic case. The interpretation of the
instability modes implies that the ellipsoidal membranes
generically decay into finger-like configurations.

We now make a quick review of the formalism for the bosonic sector
of the theory relevant to the present work. After fixing the gauge
and using reparametrization invariance of the Nambu-Gotto
Lagrangian we find that the eqs of motion for the $9$ bosonic
coordinates $X_{i}(t,\sigma_{1},\sigma_{2})$, $i=1,2,... ,9$ in
the light cone frame are: \beq \label{E1}
\ddot{X}_{i}\;=\;\{X_{k},\{X_{k},X_{i}\}\}
 \eeq
  where the Poisson bracket of two functions , $f$ and $g$ on $S^{2}$
  is defined as
 \beq \label{E2}
\{f,g\}\;=\; \frac{\partial {f}}{\partial{\cos \theta}}
\frac{\partial {g}}{\partial{ \phi}} \;-\;\frac{\partial
{f}}{\partial{\phi}}\frac{\partial {g}}{\partial{\cos\theta}} \eeq
and the remaining area preserving symmetry generated by the
constraint
 \beq\label{E3}
\left\{ X_{i}, {\dot X}_{i}\right\}= 0  \eeq In the matrix model
the above coordinates are replaced by $N \times N$ Hermitian and
traceless matrices and the corresponding equations of motion and
constraint are found by exchanging Poisson brackets with
commutators.

The first connection between the $SU(N)$ Susy Yang-Mills
truncation of the supermembrane with the recent nonperturbative
studies of string theories was discovered by Witten \cite{W96}
representing the Yang-Mills mechanics as a low energy effective
theory of bound states of $N$ $D_{0}$ branes. The $D_{0}$ branes
carry RR charge. Now it is understood how to couple the $SU(N)$
matrix model with weak background fields either directly using
supergravity arguments or truncating supermembrane Lagrangians in
weak background fields\cite{RCM}. There is an expectation that
taking appropriate limits of $N\rightarrow\infty$ for special
bound states of $N$ $D_{0}$ branes one could recover the
supermembrane or its magnetic dual, the super-five
brane\cite{PT,t01}. In the next section we turn our attention to
the analysis of the stability properties of specific classical
solutions which are spherical rotating membranes. Recent work in
the matrix model presented such a time dependent solution
representing a bound system of $N$ $D_{0}/D_{2}$ branes.

\section{Stability}

The equation of motion for the supermembrane in six dimensions may
be written as

\begin{equation}
  \label{eq:motion}
  \ddot{X}_i= \left\{ X_j,\left\{X_j, X_i \right\}\right\}
\end{equation}
where summation is implied in the j indices and $\left\{ \right\}$
stands for the Poisson bracket with respect to the angular
coordinates $\theta, \; \phi$. The Gauss constraint that also
needs to be satisfied is
\begin{equation}
  \label{eq:gauss}
 \left\{\dot{X}_i, X_i \right\}=0
\end{equation}
where $i,j=1,2..6$. We now define $Y_i\equiv X_{i+3}$ with
$i=1,2,3$. This constraint is preserved by the equations of motion
and therefore if it is initially obeyed (as is the case in what
follows) it will be obeyed at all times. The equations of motion
are \bea
  \label{eq:motion1}
  \ddot{X}_i &=& \left\{ X_j,\left\{X_j, X_i \right\}\right\}+
   \left\{ Y_j,\left\{Y_j, X_i \right\}\right\}\nn \\
  \ddot{Y}_i &=& \left\{ X_j,\left\{X_j, Y_i \right\}\right\}+
   \left\{ Y_j,\left\{Y_j, Y_i \right\}\right\}
\eea
 We now use the ansatz of a rotating spherical membrane in analogy
 with the matrix membrane ansatz  given in \cite{HS}:
\bea\label{anz} X_i &=& r_i (t) e_i(\theta, \phi)\nn \\ Y_i &=&
s_i (t) e_i(\theta, \phi) \eea where the generators
$e_i(\theta,\phi)$ are defined as \bea e_1&=&\sin\theta \;
\sin\phi \nn \\ e_2&=&\sin\theta \; \sin\phi \\ e_3&=&\cos\theta
\nn \eea
 satisfy
the relations \be \left\{ e_i,e_j \right\}=-\epsilon_{ijk} e_k \ee
Using now the ansatz (\ref{anz}) in the equations of motion
(\ref{eq:motion1}) we obtain the differential equations obeyed by
the functions $r(t)$, $s(t)$
\begin{eqnarray}
  \label{eq:s-r}
  \ddot{r}_i&=&-(r^2+s^2-r_i^2-s_i^2)r_i \\
  \ddot{s}_i&=&-(r^2+s^2-r_i^2-s_i^2)s_i
\end{eqnarray}
where $r^2=r_1^2 +r_2^2 + r_3^2$, $s^2=s_1^2+s_2^2+s_3^2$. A
particular class of solutions of (\ref{eq:s-r}) is of the form
 \begin{eqnarray}
  \label{eq:s-r-sol}
r_i &=& R_i \cos(\omega_i t + \phi_i) \\ s_i &=& R_i \sin(\omega_i
t + \phi_i)
\end{eqnarray}
with
\be
\omega_i^2 = R^2 - R_i^2 \ee where $R^2 = R_1^2 + R_2^2 + R_3^2$.

We observe that all the relations we obtained for the ansatz
(\ref{anz}) are identical with those of ref.\cite{HS}  for the
matrix model solution of a bound state of $N D_{0}/D_{2}$-branes
where the three functions $ e _{i}(\theta,\phi)$ are replaced by
N-dimensional representational matrices $
J_{\imath}(\imath=1,2,3)$ of $SU(2)$. This unique isomorphism is
due to the existence of an $SU(2)$ subgroup of the infinite
dimensional area preserving group of the sphere ($sDiff(S^2)$). It
is known that there is no other finite dimensional subalgebra of
$(sDiff(S^{2})$. As we shall see the stability analysis of the
anisotropic ($R_i \neq R_j$) membrane solution follows an
isomorphic pattern with the matrix model solution . We point out
that in ref\cite{HS} the anisotropic ($R_i \neq R_j$) matrix
solution was found to be stable under a restricted set of the
$l=1$ perturbations. In the following we extend their analysis for
every value of $l$ and we complete also the case $l=1$. The
variational equations that correspond to the splitting in eq.
$(4.3)$ between $X_{i}$ and $Y_{i}$  are: \bea \label{var} {
\ddot\delta X_{i}} \; &=& \; \left\{ \delta X_{j}, \left\{X_{j},
X_{i}\right\}\right\} + \left\{ X_j ,\left\{\delta X_j,
X_i\right\}\right\} \nn \\  &+& \left\{ X_j, \left\{ X_j, \delta
X_i \right\}\right\} + \left\{ \delta
Y_j,\left\{Y_j,X_i\right\}\right\}\nn \\ &+&
\left\{Y_j,\left\{\delta Y_j,X_i \right\}\right\} + \left\{
Y_j,\left\{ Y_j,\delta X_i\right\}\right\}\eea The corresponding
perturbation for $\delta Y_i$s and $ Y_i$s satisfy equations that
are obtained by exchanging  $ \delta X_i \leftrightarrow \delta
Y_i , X_i \leftrightarrow Y_i$ in eq(\ref{var}). The equations of
motion imply the validity of the constraint at all times \beq
\label{constr}\{ \dot{X}_i, X_i\} + \{ \dot{Y}_i , Y_i \} =0 \eeq
This is obtained by taking the time derivative of
eq.(\ref{constr}) and by applying the equations of motion and the
Jacobi identity. By expanding a configuration which at $t=0$ is
consistent with the constraint (\ref{constr}) around any classical
solution we see (by using only the linearized eqs.(\ref{var}))
that the variation $\delta X_i$ and $\delta Y_i$ satisfy the
constraint \beq \label{con1} \{ \delta \dot{X}_i, X_i\} + \{
\dot{X}_i ,\delta X_i \} +\{{\delta {\dot Y}}_i, Y_i\}+ \{{\dot
Y}_i, \delta Y_i\} = 0 \eeq for all times.

 In order to study the stability of this solution we
consider the following general form of perturbations \bea
\label{pertanz} \delta X_i (t)&=&\sum_{l,m} \epsilon_i^{lm}(t)
Y_{lm}(\theta,\phi) \nn \\ \delta Y_i (t)&=&\sum_{l,m}
\zeta_i^{lm}(t) Y_{lm}(\theta,\phi) \eea We now use the fact that
\be \left\{ e_i,Y_{lm}(\theta, \phi) \right\}=i {\hat L}_i
Y_{lm}(\theta, \phi) \ee where ${\hat L}_i$ is the angular
momentum differential operator in spherical coordinates. This
implies that \beq \left\{ e_i,Y_{lm}(\theta, \phi) \right\} =
\sum_{m'} a_{lm'}Y_{lm'}(\theta, \phi) = i \sum_{m'}
(L_i)_{mm^{'}} Y_{lm^{'}}\eeq where $L_{\imath}$ are the angular
momenta in the representation $l=(N-1)/2$. A crucial observation
is that the sum involves spherical harmonics of the same $l$ as
the spherical harmonic in the Poisson bracket. This decouples the
various $l$ fluctuation modes and simplifies the differential
equations obeyed by the modes $\epsilon_i^{lm}$ and $\zeta_i^{lm}$
as well as the gauge constraint.  This feature is specific to the
particular background solution of the spherical membrane.

The equations obeyed by the fluctuation modes $\epsilon$ and
$\zeta$ may be written as \bea \label{ezeq} {\ddot \epsilon}_i +
L_R^2\epsilon_i &=& \cos\omega_i t T_{ij}[\epsilon_j \;
cos\omega_j t + \zeta_j \; sin\omega_j t]\nn \\
 {\ddot \zeta}_i +  L_R^2\zeta_i &=&  \sin\omega_i t T_{ij}[\epsilon_j \; cos\omega_j
t + \zeta_j \; sin\omega_j t] \eea where repeated indices are
summed and the $3(2l+1) \times 3(2l+1)$ matrix $T_{ij}$ is defined
as \be \label{tdef} T_{ij}=R_i R_j(L_i L_j -2 i \epsilon_{ijk}
L_k) \ee $L_i$ is the angular momentum operator and $L_R^2 \equiv
R_1^2 L_1^2 + R_2^2 L_2^2 + R_3^2 L_3^2 $.

We now perform a rotation and define the new variables $\theta_i$
and $\eta_i$ \bea \theta_i &\equiv & \epsilon_i \cos\omega_i t +
\zeta_i \sin\omega_i t \\ \eta_i &\equiv & -\epsilon_i
\sin\omega_i t + \zeta_i \cos\omega_i t \eea

The equations obeyed by the $3(2l+1)$ vectors $\Theta=(\theta_i),
\; H=(\eta_i)$  may now be shown to be  \bea \label{syst1} {\ddot
\Theta} -2\Omega {\dot H} + [L_R^2- \Omega^2]\Theta -  T \Theta
&=& 0
\\ {\ddot H} +2\Omega {\dot \Theta} + [L_R^2-
\Omega^2]H &=& 0  \nn \eea where $\Omega=(\omega_i \delta_{ij})$,
$T=(T_{ij})$ and from the equation of motion of the background
solution we have $\omega_1^2 =  R_2^2 + R_3^2$ (with all cyclic
rotations of indices).

Perturbations of the classical solutions along the $7,8,9$
dimensions can be parametrized as \be \delta Z_i \equiv \delta
X_{i+6} \ee with $i=1,2,3$ and \be  \ddot{\delta Z_i}= \left\{
X_j,\left\{X_j, \delta Z_i \right\}\right\} + \left\{
Y_j,\left\{Y_j, \delta Z_i \right\}\right\} \ee With the
definition \be \delta Z_i (t)=\sum_{l,m} q_i^{lm}(t)
Y_{lm}(\theta,\phi) \ee we obtain \be {\ddot q_i} = - L_R^2 \; q_i
\ee which implies stability along the 7,8,9 dimensions since
$L_R^2$ is positive definite.

In order to study the stability of the system (\ref{syst1}) we
must convert it to an eigenvalue problem. Assuming $\Theta =
e^{i\lambda t} a$ and $H=e^{i \lambda t} b$ we obtain \bea Ma &=&
\lambda^2 a + 2i\lambda \Omega b \\ (L_R^2 - \Omega^2)b &=&
\lambda^2 b - 2i\lambda \Omega a \eea This is a complicated
$6(2l+1)$ dimensional quadratic eigenvalue problem for which there
is no standard mathematical theory. It can be formulated as a
non-commutative quadratic equation (with matrix
coefficients)\cite{zs00}. In what follows we shall bypass this
problem applying a method introduced by N. Papanicolaou\cite{pap}
and which transforms the problem to a linear eigenvalue one.

 We define the new variables $\Theta_1$, $\Theta_2$, $H_1$ and $H_2$
as follows: \bea {\dot \Theta_1} &=& \Theta_2 \\ {\dot H_1} &=&
H_2 \eea  Thus the system (\ref{syst1}) may be written as \bea
 {\dot \Theta_2} -2\Omega H_2 + [L_R^2-
\Omega^2]\Theta_1 -  T\Theta_1 &=& 0 \\ {\dot H_2} +2\Omega
\Theta_2 + [L_R^2- \Omega_2]H_1 &=& 0 \eea or

\begin{equation}
\label{system} {\dot X} =   \left(
\begin{array}{cccc} 0 & 0 & 1 & 0\\0 & 0 & 0 & 1 \\
-(L_R^2 - \Omega^2 - T) & 0 & 0 & 2\Omega \\ 0 & -(L_R^2 -
\Omega^2) & -2 \Omega& 0 \end{array}\right) X
\end{equation}
 where
\begin{equation}
\label{xdef} X\equiv \left( \begin{array}{c} \Theta_1
\\ H_1
\\ \Theta_2 \\ H_2
\end{array} \right) = e^{i \lambda t}\left( \begin{array}{c} a_1
\\ b_1
\\ a_2 \\ b_2
\end{array} \right)= e^{i \lambda t} A
\end{equation}

Thus we are led to the eigenvalue problem \be \label{egv} i
\lambda \; A = M \; A \ee where $M$ is defined by eq.
(\ref{system}). The membrane solution is stable if and only if all
the eigenvalues $\lambda$ are real. The matrix $M$ is a well
defined matrix of dimension $12(2l+1)\times 12(2l+1)$ and
therefore it is straightforward to solve the eigenvalue problem
(\ref{egv}) numerically for any $l$.

The stability analysis for the isotropic case was performed in ref
\cite{afp00,ssnew} and the eigenvalues along with their
degeneracies are shown in Table I

In this case it is obvious that all $\lambda_i^2$ are non-negative 
and therefore the eigenfrequencies $\lambda_i$ are all real. This 
implies that the isotropic membrane solution studied is stable to 
first order in perturbation theory. 

The $2l+1$ zero modes are due to gauge degrees of freedom and 
survive in the anisotropic case. For the gauge zero modes the 
corresponding constant vectors $\theta_i$, $\eta_i$ are 
$\theta_i=R_i L_i v$, $\eta_i=0$ for any $2l+1$ dimensional vector 
$v$. These modes satisfy the equation \be \label{modsat} (L_R^2 - 
\Omega^2 -T)_{ij} \theta_j =0 \ee The physical modes must satisfy 
the constraint equation (\ref{con1}) which may be written as \be 
R_iL_i (2\omega_i \eta_i -{\dot \theta}_i) =0 \label{conx} \ee 
where summation is implied in $i$. 

We have checked numerically the expected orthogonality between the 
physical modes and the gauge zero modes. The modes orthogonal to 
the gauge zero modes satisfy the constraint equations ie are 
physical. 
\begin{table} \caption{The
eigenvalues along with their degeneracy for the isotropic membrane 
stability problem.} 
\begin{tabular}{ccc}
Degeneracy&$\lambda_a^2$&$\lambda_b^2$\\ \tableline $2l+1$ & 0 & 
$l^2 + l + 6$  \\ \tableline $2l+3$&$l^2 - 3l + 2$ & $l^2 + 3l + 
2$\\ \tableline $2l-1$&$l^2 - l$ & $l^2 + 5l + 6$ \\ 
\end{tabular}
\end{table}
\begin{table}
\caption{The maximum imaginary eigenvalue for various degrees of 
anisotropy (we have set $R_1=R_2=1$ and 
$R_3=1-\epsilon$).\label{table2}} 
\begin{tabular}{ccccc}
 &$\epsilon = 1$&$\epsilon = 0.9$&$\epsilon = 0.4$&$\epsilon
= 0.1$ \\ \tableline $l=1$ & 0 & 0& 0& 0 
\\ \tableline $l=2$&0& 0.44 & 0.92 & 0.99\\ \tableline
$l=3$& 0 & 0 & 1.16 & 1.4 \\ 
\end{tabular}
\end{table}
This can be proven by taking the inner product of the left hand 
side of the constraint equation (\ref{conx}) with an arbitrary 
$2l+1$ dimensional vector $v$ and using the properties of the 
gauge zero modes discussed above.  

In the general complete analysis of the anisotropic membrane, for 
$l=1$, we find that all modes are stable in agreement with ref 
\cite{ssnew} where only some of the $l=1$ modes were studied. 
However we have found unstable modes for $l \geq 2$ and therefore 
we conclude that the membrane configuration is unstable in this 
case. In particular, we have found that the instability for $l=2$ 
persists for all values of non-zero anisotropy.  However, for $l> 
2$ there is a minimum anisotropy below which we have stability of 
the corresponding modes. In Table II we show the imaginary part of 
the most unstable eigenvalue for various $l$ values as a function 
of the anisotropy while in Fig. 1 we show the dependence of the 
most unstable zero modes on the anisotropy. 
\begin{figure}[htbp]
  \begin{center}
    \centerline{\epsfxsize=9cm
    \epsfbox{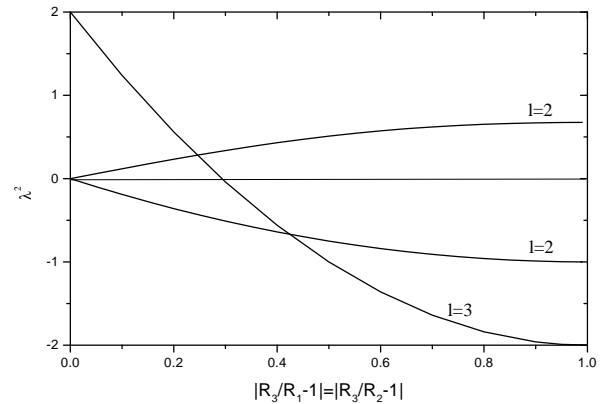}}
    \caption{The instability split of the zero mode leads to instability (imaginary eigenvalue $\lambda$)
    for any value of the anisotropy for $l=2$. For $l=3$ the anisotropy appears for a finite value of the anisotropy.}
    \label{fig:1}
  \end{center}
\end{figure}
%\begin{figure}[htbp]
%  \begin{center}
%    \centerline{\epsfxsize=9cm
%    \epsfbox{fig2.ps}}
%    \caption{The imaginary part of the most unstable eigenvalue for $l=2$ as a function of the anisotropy.}
%    \label{fig:2}
%  \end{center}
%\end{figure}
%In Fig. 2 we show the imaginary part of the most unstable mode
%eigenvalue as a function of the anisotropy $R_1/R_2 = R_1/R_3$ for
%$l=2$. Clearly the imaginary part of the most unstable eigenvalue
%approaches an asymptotic value which for $l=2$ is equal to 2.
%We have also found that for large values of the anisotropy there
%is only one unstable mode.

The derived instability of the anisotropic membrane is reminiscent
of the Bohr atom where the classical electron spirals towards the
nucleus emitting electromagnetic radiation. Thus the stability
analysis we have performed has implications for the semiclassical
quantization of the supermembrane indicating that there is an
additional principle required in order to stabilize quantum
mechanically the rotating anisotropic membranes against emission
of spikes and D0 branes. In a later publication we will present a
detailed description of the findings presented here.

\section{Acknowledgements}
We are grateful to N. Papanicolaou and to G. K. and K. G.  Savvidy
for useful discussions. This work was supported by the EU under
the TMR No CT2000-00122,00431.

\bibliographystyle{prsty}

\bibliography{bibliog}

\end{document}